# Emergent *d*-wave altermagnetism in orthogonally twisted bilayer CrPS$_4$

Alberto M. Ruiz[1,||], Diego López-Alcalá[1,||], Rafael González-Hernández[2], José J. Baldoví[1,*]

[1]Instituto de Ciencia Molecular, Universitat de València, Catedrático José Beltrán 2, 46980 Paterna, Spain

[2]Departamento de Física y Geociencias, Universidad del Norte, Barranquilla, Colombia

[||]These authors have contributed equally to this work

Email: * j.jaime.baldovi@uv.es

KEYWORDS: Altermagnetism, twistronics, CrPS$_4$, first-principles calculations.

Twistronics is a powerful strategy to engineer novel quantum states by controlling the relative orientation between layered materials. Here, we demonstrate that an orthogonally twisted bilayer CrPS$_4$ shows *d*-wave altermagnetism driven purely by structural rotation. Symmetry analysis reveals that the twisted stacking breaks partial translational combined with time-reversal symmetry, leading to a fourfold rotation relation between opposite spin sublattices, enabling altermagnetism. First-principles calculations demonstrate a sizable non-relativistic spin splitting of up to 68 meV around the Fermi level. We further show that the altermagnetic state can be further stabilized through interlayer compression and modulation of the on-site Coulomb interaction. The resulting band structure exhibits pronounced spin-dependent anisotropy, enabling efficient spin to charge conversion reaching ~50% near the Fermi level and sizable giant magnetoresistance. These results establish twisted CrPS$_4$ as a realistic platform for altermagnetism and highlights twistronics as a versatile route for advanced spintronics applications.

Twistronics— the controlled rotational alignment of two-dimensional (2D) materials—has emerged as a powerful paradigm for engineering novel quantum states, leading to phenomena such as moiré flat bands, unconventional superconductivity or ferroelectricity.[1–7] In magnetic van der Waal (vdW) materials, twisting allows the modification of the symmetry relations between spin sublattices without chemical substitution, influencing their magnetic response. For instance, in CrI$_3$, small twist angles induce an interplay between monoclinic and rhombohedral stacking configurations associated with interlayer antiferromagnetic (AF) and ferromagnetic (FM) states, respectively.[8–11] The resulting competition between stacking-dependent exchange and magnetic anisotropy generate noncollinear topological spin textures and theoretically predicted spin–orbit coupling (SOC)–driven local ferroelectric polarization.[12–16]

Within this broader magnetic landscape, altermagnetism has recently emerged as a magnetic state characterized by a momentum-dependent spin splitting in the absence of net magnetization.[17–20] Therefore, altermagnets combine key advantages of both FM and AF materials such as absence of stray fields, robustness against external perturbations, and ultrafast spin dynamics, while enabling sizable spin-polarized electronic responses.[21–23] In altermagnets, spin splitting originates from crystal symmetry rather than SOC and exhibits characteristic patterns in momentum space. Those can be classified in *d*-, *g*-, or *i*-wave type corresponding to 2, 4 or 6 spin degenerate nodal surfaces crossing Γ point, respectively.[24,25]

Twisting layered antiferromagnets has recently been proposed as an effective route for altermagnetism in 2D systems.[26–29] Most reported candidates to date rely on relatively small twisting angles between layers, resulting in higher-order *i*-wave spin-splitting patterns. Consequently, their symmetry enforces multiple sign changes of the spin polarization, often leading to equivalent group velocities for opposite spin channels and suppressing spin-polarized currents. In contrast, *d*-wave altermagnets are particularly desirable, as their reduced number of spin-degenerate nodal surfaces results in anisotropic group velocities that enable spin-polarized transport.[30–33] In this context, a 90° rotation between adjacent layers provides a natural route to realize *d*-wave altermagnetism by generating two orthogonal spin channels with opposite spin polarization.[26,34]

Recently, 90º twisted bilayer CrSBr has been theoretically proposed as a *d*-wave altermagnet based on the AF interlayer coupling of the pristine system and the experimental realization of the orthogonally stacked architecture.[26,35,36] However, its in-plane anisotropy results in two decoupled layers in the rotated form, where spins are independently aligned along their respective easy axes.[36,37] This behaviour, combined with its extremely weak interlayer antiferromagnetism in the pristine phase, precludes altermagnetism.[38–40] In contrast, CrPS$_4$ emerges as a more robust and promising platform. CrPS$_4$ is a vdW A-type AF semiconductor and, importantly, exhibits out-of-plane magnetic anisotropy together with significantly stronger interlayer AF coupling.[41–43] These characteristics suggest that interlayer antiferromagnetism may be preserved under rotational stacking, providing an ideal platform for altermagnetism.

In this work, we demonstrate that orthogonally twisted bilayer CrPS$_4$ is a $d$-wave altermagnet by combining first-principles calculations and symmetry analysis. In particular, we show that a 90° rotation between adjacent layers leads to opposite spin sublattices related by fourfold rotation along the z direction. This configuration stabilizes a $d$-wave AM state, giving rise to a momentum-dependent spin-splitting around the Fermi level of 68 meV even in the absence of SOC. Moreover, the twisted structure enables efficient spin to charge conversion near the Fermi level, reaching values approaching 50%. Our results stablish twisted CrPS$_4$ as an experimentally accessible candidate for altermagnetism.

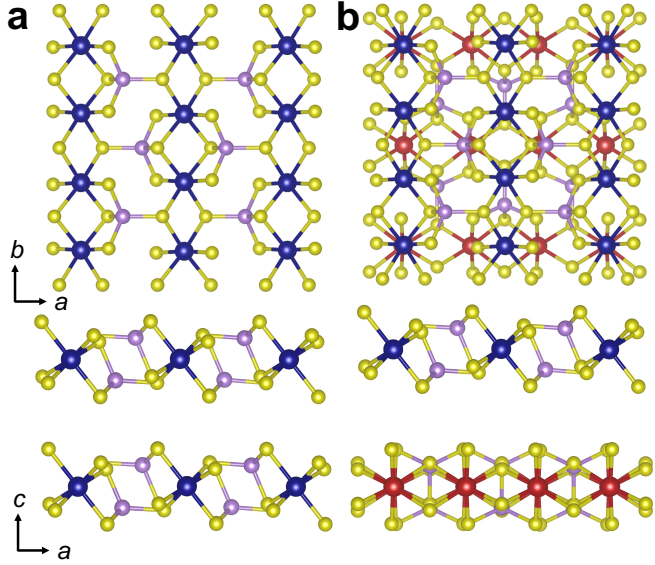

**Figure 1**. (a) Top and side views of pristine bilayer and (b) 90° twisted bilayer CrPS$_4$. Color code: Cr (blue in the pristine bilayer; blue and red in the twisted bilayer for clarification), P (pink), and S (yellow).

An illustration of bilayer CrPS$_4$ is shown in Figure 1a. Each layer consists of edge-sharing CrS$_6$ octahedra and PS$_4$ tetrahedra extended along the $b$ and $a$ directions, respectively, resulting in pronounced in-plane structural anisotropy.[44,45] To characterize the electronic and magnetic properties of CrPS$_4$, we perform first-principles calculations within the DFT+U framework. The optimized lattice parameters are $a$ = 10.88 Å and $b$ = 7.28 Å, showing an interlayer separation of 2.52 Å.[44] The interlayer exchange coupling ($J_{int}$) is evaluated from the energy difference between AF and FM configurations. Our results yield $J_{int}$ = -0.59 meV/ Cr, correctly capturing the AF ground state of the material.[46,47] This value is obtained using a Hubbard parameter of U = 1.6 eV, which is selected since it reproduces the magnitude of $J_{int}$ extracted from experimental findings and is in good agreement with the one obtained using HSE06 hybrid functionals (Figure S1).[47] Therefore, we adopt this U = 1.6 eV for our subsequent analysis.

The spin-symmetry operations [$R_i$||$R_j$] for bilayer CrPS$_4$ are defined as follows. The operation [$C_2$||$t$] connects the two layers through AA stacking, where the transformation on the left (right) of the double vertical bar acts in spin (real crystallographic) space. Here, $C_2$ corresponds to a spin inversion operation, while $t$ denotes a conventional lattice translation. This symmetry relation imposes conventional AF order and leads to a spin degenerate band structure (Figure S2).[42,46] Furthermore, CrPS$_4$ exhibits out-of-plane magnetic anisotropy arising from the combined effects of spin–orbit coupling (SOC) and dipole–dipole interactions.[42] Including both contributions, we obtain $MAE_{bc}$ = $E_b$ – $E_c$ = 24.5 μeV/Cr and $MAE_{ac}$ = $E_a$ – $E_c$ = 39.8 μeV/Cr, stablishing the out-of-plane direction as the easy magnetization axis (Table S1).[41,42,46]

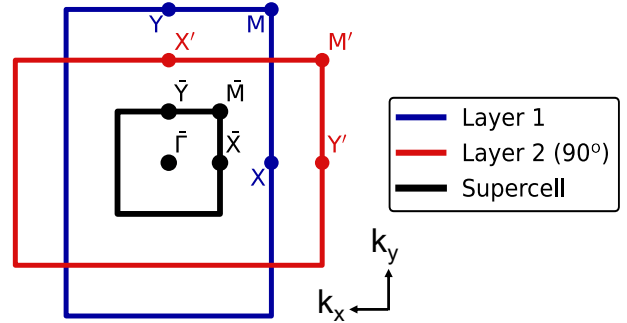

**Figure 2**. Schematic representation of the Brillouin zones of the untwisted layer (Layer 1), the twisted layer (Layer 2), and the common supercell employed for the 90° twisted CrPS$_4$ heterostructure.

Then, we investigate a twisted bilayer configuration upon a rotation angle of 90°. To construct such heterostructure, we create a 2×3 supercell with lattice parameters $a$ = 21.76 Å and $b$ = 21.84 Å. A small uniform strain is then applied to both layers to achieve commensurability, resulting in a square lattice with $a$ = $b$ = 21.80 Å, with a negligible applied strain of 0.18% for each layer (Table S2). Upon rotation, the Brillouin zone of the twisted layer is correspondingly rotated (Figure 2). Specifically, the Γ–X direction along $k_x$ transforms into Γ–X' along $k_y$, while Γ–Y rotates toward the x axis, which we denote as Γ–Y'. In the common supercell, the square lattice leads to two equivalent reciprocal lattice vectors lengths, denoted as $\bar{\Gamma}$–$\bar{X}$ and $\bar{\Gamma}$–$\bar{Y}$.

The orthogonal twist breaks the symmetry associated with the [$C_2$||$t$] operation. Instead, opposite spin sublattices are connected through 90° spatial rotations, described by the [$C_2$||$C_4$] operation, which generates the symmetry conditions required for altermagnetism. The calculated electronic band structure exhibits clear momentum-dependent spin splitting along $\bar{\Gamma}$–$\bar{X}$ and $\bar{\Gamma}$–$\bar{Y}$ (Figure 3a). For the rotated layer, the sequence of spin channels is inverted, reflecting the 90° rotation between those. The intrinsic band gap of the pristine bilayer is preserved upon twisting, showing a value of 1.2 eV.

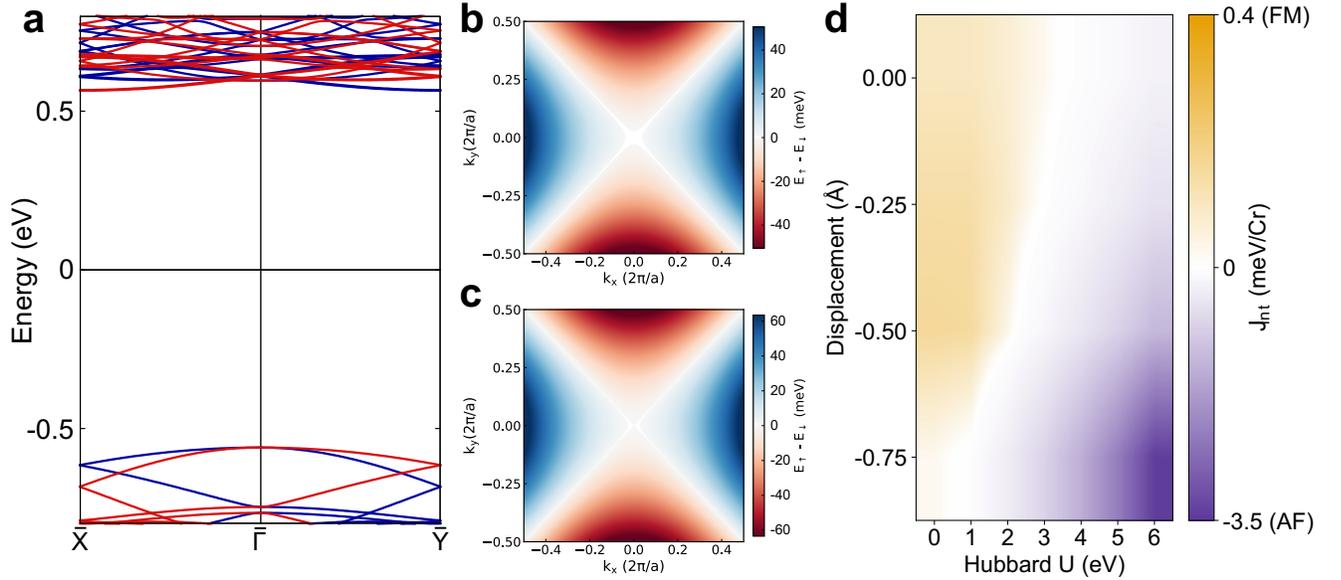

**Figure 3**. (a) Electronic band structure of twisted 90° bilayer $CrPS_4$, together with the corresponding 2D maps of spin splitting for the (b) conduction band minimum and (c) valence band maximum. Blue and red color in the band structure denotes spin up and spin down states, respectively. (d) Variation of the interlayer exchange coupling $J_{int}$ as a function of the on-site Hubbard U parameter and interlayer displacement. A displacement of 0.00 Å corresponds to the equilibrium interlayer distance of twisted bilayer $CrPS_4$ (3.25 Å).

The electronic band structure exhibits pronounced spin splitting, which is larger in the valence band maximum than in the conduction band minimum. From the 2D spin-splitting maps (Figures 3b, c), one can observe that the splitting vanishes at $\bar{\Gamma}$ and increases continuously toward the $\bar{\Gamma}$–$\bar{X}$ and $\bar{\Gamma}$–$\bar{Y}$ directions, reaching a maximum value of ±68 meV at $\bar{X}$ and $\bar{Y}$, respectively. A similar trend is observed in the conduction band, with a maximum splitting of ±43 meV. In both cases, the momentum dependence of the splitting follows a characteristic d-wave symmetry, where its sign is reverted every 90°. It is also observed that a symmetry-protected nodal line emerges along the directions $\bar{\Gamma}$–$\bar{M}$. The inclusion of SOC effects leaves the band structure essentially unchanged (Figure S3), confirming the non-relativistic origin of the splitting. Upon rotation, the preferential out-of-plane spin orientation is preserved, with a MAE of 22.3 μeV/Cr required to rotate spins toward the in-plane direction (Table S3).

Twisted bilayer $CrPS_4$ exhibits an increased interlayer separation of 3.25 Å compared to 2.50 Å in the untwisted structure.[48] This larger distance reduces the interlayer coupling, leading to a weak FM state with $J_{int}$ = 0.085 meV/Cr. Owing to the vdW nature of $CrPS_4$, this interaction can be effectively modified through uniaxial pressure applied on the out-of-plane direction. Indeed, experiments on bulk $CrPS_4$ show that lattice compression decreases the interlayer spacing and enhances the AF character of $J_{int}$ by 45%.[49] A similar trend has been experimentally reported for CrSBr, where a ~6% reduction of the interlayer distance enhances the AF coupling by 400%,[50] reaching values of ~1700% at higher pressure levels.[51] Therefore, reducing the layer separation in twisted $CrPS_4$ is expected to stabilize an AF ground state, providing a realistic route to experimentally access the AM phase.

To quantify this effect, we compute the evolution of $J_{int}$ as a function of the interlayer distance for U = 1.6 eV (Figure 3d). Upon compression, $J_{int}$ evolves continuously toward an AF regime, reaching -0.18 meV/Cr at a displacement of −0.75 Å, which corresponds to the interlayer separation of the pristine bilayer. Notably, this value is smaller compared to the untwisted case ($J_{int}$ = -0.59 meV/Cr), despite the identical vertical spacing. This difference originates from their different stacking geometry. In the pristine structure, Cr atoms are vertically aligned between adjacent layers (Figure 1a), maximizing the AF coupling. On the other hand, the twisted configuration induces a reduced orbital overlap between interface atoms (Figure 1b), which weakens the AF interaction. While $J_{int}$ is highly sensitive to interlayer variation, the magnetic anisotropy remains largely unaffected. Even under a maximum compression of -0.75 Å, the system preserves an out-of-plane anisotropy (Figure S4), thereby retaining altermagnetism.

Apart from the vertical displacement between adjacent layers, the magnetic properties of 2D materials can be significantly influenced by the effective on-site Coulomb interaction U, where its magnitude can be controlled via the dielectric environment or external fields.[52–54] In particular, stronger environmental screening (e.g., high-dielectric substrates) reduces U, whereas weaker screening–such as low-dielectric substrates or suspended configurations– enhances it.[55,56] Additionally, electrostatic gating can modify screening through gate-induced carriers, enabling a continuous renormalization of U, as demonstrated in $CrBr_3$.[57] Therefore, we explore the stabilization of the AM ground state upon variation of Hubbard U parameter.

We find that increasing (decreasing) U leads to a continuous stabilization (reduction) of the AF interlayer coupling. At the equilibrium interlayer distance (displacement = 0 Å), $J_{int}$ evolves from a weak FM value of 0.085 meV/ Cr at U = 1.6 eV to an enhanced AF character with $J_{int}$ = -0.23 meV/Cr at U = 6 eV. This trend persists across all interlayer displacements, indicating that the combined effect of increasing U and layer compression progressively enhances the AF interaction, reaching $J_{int}$ = -3.42 meV/Cr for U = 6 eV and a vertical displacement of -0.75 Å. Despite the substantial modification of $J_{int}$, the out-of-plane spin orientation remains essentially unchanged (Figure S4), while the maximum spin-splitting decreases as one increases the Hubbard U and decreases the interlayer displacement (Figure S5).

In altermagnets, the anisotropic spin-momentum coupling leads to spin-polarized bands with direction-dependent group velocities, resulting in distinct conductivities for opposite spin channels. Consequently, the electrical transport becomes anisotropic and spin dependent. Figure 4a shows the electrical conductivity components $\sigma_{xx}$, $\sigma_{xy}$, and $\sigma_{yy}$ resolved by spin channel. A suppression of conductivity is observed around the Fermi level ($E_f$), indicating a null carrier density regime consistent with a semiconducting gap. Away from $E_f$, the longitudinal components $\sigma_{xx}$ and $\sigma_{yy}$ increase rapidly and reach values on the order of $10^4$ S/cm. The longitudinal conductivities exhibit a spin-dependent anisotropy governed by $C_4$ symmetry, enforcing $\sigma_{xx}^\uparrow = \sigma_{yy}^\downarrow$ and vice versa. In contrast, the transverse conductivity $\sigma_{xy}$ vanishes across the entire energy range due to the spin-degenerate bands along the nodal $\bar{\Gamma}$–$\bar{M}$ direction (Figures 2b and c).

This symmetry-driven anisotropy has consequences for spin transport, as reflected from the calculation of the spin to charge efficiency (SCE), which is defined as:

$$\text{SCE} = \frac{\sigma_{ii}^\uparrow - \sigma_{ii}^\downarrow}{\sigma_{ii}^\uparrow + \sigma_{ii}^\downarrow} \quad (1)$$

Here, $ii$ corresponds to either the $xx$ or $yy$ components and SCE gives the ratio between the spin conductivity and charge conductivity.

Figure 4b shows that spin polarization exhibits opposite signs for the longitudinal $xx$ and $yy$ components, reaching magnitudes of up to 50% below the $E_f$. This indicates that charge currents flowing along orthogonal directions are carried by opposite spin channels. As a result, the system supports the generation of a spin current without requiring net magnetization, a hallmark of $d$-wave altermagnetism.[58] Above $E_f$, the SCE reaches 25%, comparable to values reported for $RuO_2$ (27 %)[59] and $V_2Te_2O$ (32 %),[60] and higher than the one for $Mn_3Sn$ (15 %).[61] Notably, magnetotransport measurements on $CrPS_4$ indicate that the Fermi level is located close to the conduction band edge, typical of a n-type semiconductor.[46,62] Therefore, slight electron doping can populate the conduction band edge and enable a substantial SCE.

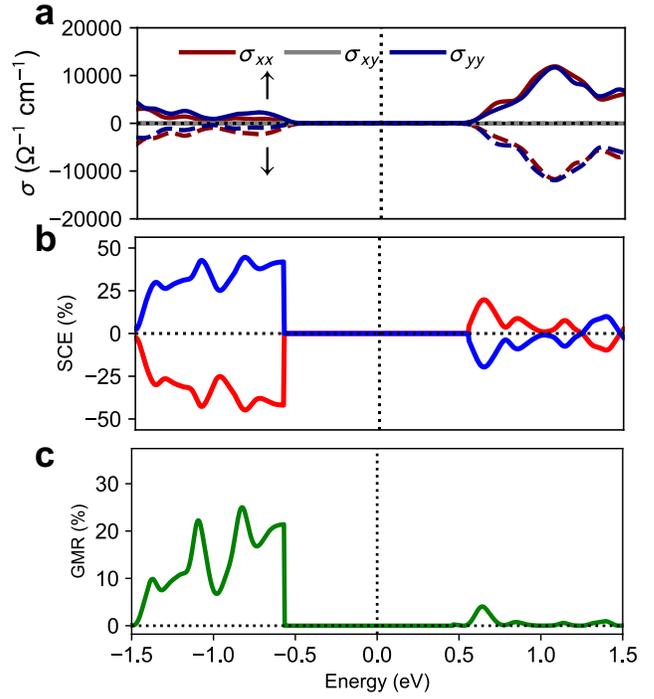

**Figure 4**. (a) Longitudinal ($\sigma_{xx}$ and $\sigma_{yy}$) and transverse ($\sigma_{xy}$) conductivities. The solid (dashed) lines correspond to the spin up (down) contributions, with the latter plotted with negative sign to enable comparison. b) SCE and c) GMR for twisted $CrPS_4$.

Finally, Figure 4c shows the giant magnetoresistance (GMR), which is derived from the ratio of spin-resolved conductivities for both spin channels, $R_\sigma = \sigma_{xx}^\uparrow/\sigma_{xx}^\downarrow = \sigma_{xx}^\uparrow/\sigma_{yy}^\uparrow = \sigma_{yy}^\downarrow/\sigma_{xx}^\downarrow$. Specifically, the GMR has the form[63]:

$$\text{GMR} = \frac{1}{4}\left(R_\sigma + \frac{1}{R_\sigma} - 2\right) \quad (2)$$

A significant GMR signal, reaching values up to 25%, is observed below the Fermi level, driven by the strong imbalance between $\sigma_{xx}^\uparrow$ and $\sigma_{xx}^\downarrow$. The GMR vanishes along the electronic gap, reflecting the null conductivity in this region, and remains comparatively small above at the conduction band edge where the spin asymmetry is weaker. These results demonstrate that the interplay between symmetry-enforced anisotropy and the energy-dependent electronic structure enables both efficient spin current generation and sizable magnetoresistive effects.

In conclusion, we demonstrate that orthogonally twisted bilayer $CrPS_4$ stands as a promising candidate for $d$-wave altermagnetism. Our results demonstrate that a 90º twist angle induces a $[C_2\|C_4]$ operation connecting opposite spin sublattices, generating a sizable non-relativistic spin splitting of up to 68 meV around the Fermi level. Our calculations show that the stability of the AM state can be reinforced by reducing the interlayer separation and through modification of the Coulomb screening. Additionally, the calculated band structure exhibits strong spin-dependent anisotropy, leading to a significant 50% (25%) spin to charge conversion below (above) the Fermi level and sizable giant magnetoresistance. Overall, our findings identify twisted CrPS$_4$ as an experimentally viable platform for altermagnetism and highlight twistronics as an effective route to engineer magnetism in 2D materials.

## METHODS

Spin-polarized density functional theory (DFT) calculations for CrPS$_4$ were performed using the Vienna Ab initio Simulation Package (VASP).[64] The exchange-correlation energy was treated within the generalized gradient approximation (GGA). To construct the 90º twisted heterostructure, a 2×3 supercell is employed, with the corresponding atomic positions and lattice parameters provided in Table S2. To accurately describe the electronic and magnetic properties, the DFT+U approach was employed, with an effective Hubbard parameter of U = 1.6 eV. This value of U was selected as it reproduces the experimental reported $J_{int}$ = -0.59 meV/Cr,[47] and yields results consistent with those obtained using the HSE06 hybrid functional (Figure S1). Van der Waals (vdW) interactions between adjacent CrPS$_4$ layers were treated using the DFT-D2 scheme. A comparison with the DFT-D3 method shows negligible differences, with both approaches reproducing $J_{int}$ = -0.59 meV/Cr for U = 1.6 eV (Figure S2). Boltzmann transport calculations we performed using the Wannier90 code.[65] The longitudinal and transversal conductivities were computed using a 120×120×1 crystal momentum mesh. The temperature was set to T = 300 K and the scattering rate to $\Gamma$ = 10 meV. Additional calculations were performed for $\Gamma$ = 1 meV and $\Gamma$ = 100 meV (Figures S6-8), showing that the conductivities are inversely proportional to $\Gamma$,[61] while the SCE and GMR remain robust against variations of $\Gamma$.[58]

## ACKNOWLEDGMENTS

The authors acknowledge financial support from the European Union (ERC-2021-StG101042680 2D-SMARTiES), the Spanish MCIU (PID2024-162182NA-I00 2D-MAGIC), Spanish MICINN (Excellence Unit "Maria de Maeztu" CEX2024-001467-M) and the Generalitat Valenciana (grant CIDEXG/2023/1). A.M.R. thanks the Spanish MIU (Grant No FPU21/04195). The calculations were performed on the HAWK cluster of the 2D Smart Materials Lab hosted by Servei d'Informàtica of the Universitat de València.

## REFERENCES

(1) Castellanos-Gomez, A.; Duan, X.; Fei, Z.; Gutierrez, H. R.; Huang, Y.; Huang, X.; Quereda, J.; Qian, Q.; Sutter, E.; Sutter, P. Van Der Waals Heterostructures. *Nature Reviews Methods Primers* **2022**, *2* (1), 58.

(2) Yasuda, K.; Wang, X.; Watanabe, K.; Taniguchi, T.; Jarillo-Herrero, P. Stacking-Engineered Ferroelectricity in Bilayer Boron Nitride. *Science* **2021**, *372* (6549), 1458–1462.

(3) Cao, Y.; Fatemi, V.; Fang, S.; Watanabe, K.; Taniguchi, T.; Kaxiras, E.; Jarillo-Herrero, P. Unconventional Superconductivity in Magic-Angle Graphene Superlattices. *Nature* **2018**, *556* (7699), 43–50.

(4) Cao, Y.; Fatemi, V.; Demir, A.; Fang, S.; Tomarken, S. L.; Luo, J. Y.; Sanchez-Yamagishi, J. D.; Watanabe, K.; Taniguchi, T.; Kaxiras, E.; Ashoori, R. C.; Jarillo-Herrero, P. Correlated Insulator Behaviour at Half-Filling in Magic-Angle Graphene Superlattices. *Nature* **2018**, *556* (7699), 80–84.

(5) Weston, A.; Castanon, E. G.; Enaldiev, V.; Ferreira, F.; Bhattacharjee, S.; Xu, S.; Corte-León, H.; Wu, Z.; Clark, N.; Summerfield, A.; Hashimoto, T.; Gao, Y.; Wang, W.; Hamer, M.; Read, H.; Fumagalli, L.; Kretinin, A. V.; Haigh, S. J.; Kazakova, O.; Geim, A. K.; Fal'ko, V. I.; Gorbachev, R. Interfacial Ferroelectricity in Marginally Twisted 2D Semiconductors. *Nat. Nanotechnol.* **2022**, *17* (4), 390–395.

(6) Andrei, E. Y.; Efetov, D. K.; Jarillo-Herrero, P.; MacDonald, A. H.; Mak, K. F.; Senthil, T.; Tutuc, E.; Yazdani, A.; Young, A. F. The Marvels of Moiré Materials. *Nat. Rev. Mater.* **2021**, *6* (3), 201–206.

(7) He, F.; Zhou, Y.; Ye, Z.; Cho, S.-H.; Jeong, J.; Meng, X.; Wang, Y. Moiré Patterns in 2D Materials: A Review. *ACS Nano* **2021**, *15* (4), 5944–5958.

(8) Song, T.; Sun, Q.-C.; Anderson, E.; Wang, C.; Qian, J.; Taniguchi, T.; Watanabe, K.; McGuire, M. A.; Stöhr, R.; Xiao, D.; Cao, T.; Wrachtrup, J.; Xu, X. Direct Visualization of Magnetic Domains and Moiré Magnetism in Twisted 2D Magnets. *Science* **2021**, *374* (6571), 1140–1144.

(9) Xu, Y.; Ray, A.; Shao, Y.-T.; Jiang, S.; Lee, K.; Weber, D.; Goldberger, J. E.; Watanabe, K.; Taniguchi, T.; Muller, D. A.; Mak, K. F.; Shan, J. Coexisting Ferromagnetic–Antiferromagnetic State in Twisted Bilayer CrI$_3$. *Nat. Nanotechnol.* **2022**, *17* (2), 143–147.

(10) Xie, H.; Luo, X.; Ye, G.; Ye, Z.; Ge, H.; Sung, S. H.; Rennich, E.; Yan, S.; Fu, Y.; Tian, S.; Lei, H.; Hovden, R.; Sun, K.; He, R.; Zhao, L. Twist Engineering of the Two-Dimensional Magnetism in Double Bilayer Chromium Triiodide Homostructures. *Nat. Phys.* **2022**, *18* (1), 30–36.

(11) Cheng, G.; Rahman, M. M.; Allcca, A. L.; Rustagi, A.; Liu, X.; Liu, L.; Fu, L.; Zhu, Y.; Mao, Z.; Watanabe, K.; Taniguchi, T.; Upadhyaya, P.; Chen, Y. P. Electrically Tunable Moiré Magnetism in Twisted Double Bilayers of Chromium Triiodide. *Nat. Electron.* **2023**, *6* (6), 434–442.

(12) Xie, H.; Luo, X.; Ye, Z.; Sun, Z.; Ye, G.; Sung, S. H.; Ge, H.; Yan, S.; Fu, Y.; Tian, S.; Lei, H.; Sun, K.; Hovden, R.; He, R.; Zhao, L. Evidence of Non-Collinear Spin Texture in Magnetic Moiré Superlattices. *Nat. Phys.* **2023**, *19* (8), 1150–1155.

(13) Wong, K. C.; Peng, R.; Anderson, E.; Ross, J.; Yang, B.; Cheng, M.; Jayaram, S.; Lenger, M.; Zhou, X.; Kong, Y. T.; Taniguchi, T.; Watanabe, K.; McGuire, M. A.; Stöhr, R.; Tsen, A. W.; Santos, E. J. G.; Xu, X.; Wrachtrup, J. Super-Moiré Spin Textures in Twisted Two-Dimensional Antiferromagnets. *Nat. Nanotechnol.* **2026**, *21* (3), 359–365.

(14) Xiao, F.; Chen, K.; Tong, Q. Magnetization Textures in Twisted Bilayer CrX$_3$ (X = Br, I). *Phys. Rev. Res.* **2021**, *3* (1), 013027.

(15) Akram, M.; LaBollita, H.; Dey, D.; Kapeghian, J.; Erten, O.; Botana, A. S. Moiré Skyrmions and Chiral Magnetic Phases in Twisted CrX$_3$ (X = I, Br, and Cl) Bilayers. *Nano Lett.* **2021**, *21* (15), 6633–6639.

(16) Fumega, A. O.; Lado, J. L. Moiré-Driven Multiferroic Order in Twisted CrCl$_3$, CrBr$_3$ and CrI$_3$ Bilayers. *2d Mater.* **2023**, *10* (2), 025026.

(17) Šmejkal, L.; Sinova, J.; Jungwirth, T. Beyond Conventional Ferromagnetism and Antiferromagnetism: A Phase with Nonrelativistic Spin and Crystal Rotation Symmetry. *Phys. Rev. X* **2022**, *12* (3), 031042.

(18) Šmejkal, L.; Sinova, J.; Jungwirth, T. Emerging Research Landscape of Altermagnetism. *Phys. Rev. X* **2022**, *12* (4), 040501.

(19) Mazin, I. Editorial: Altermagnetism—A New Punch Line of Fundamental Magnetism. *Phys. Rev. X* **2022**, *12* (4), 040002.

(20) Šmejkal, L.; González-Hernández, R.; Jungwirth, T.; Sinova, J. Crystal Time-Reversal Symmetry Breaking and Spontaneous Hall Effect in Collinear Antiferromagnets. *Sci. Adv.* **2020**, *6* (23), eaaz8809.

(21) Bai, L.; Feng, W.; Liu, S.; Šmejkal, L.; Mokrousov, Y.; Yao, Y. Altermagnetism: Exploring New Frontiers in Magnetism and Spintronics. *Adv. Funct. Mater.* **2024**, *34* (49), 2409327.

(22) Song, C.; Bai, H.; Zhou, Z.; Han, L.; Reichlova, H.; Dil, J. H.; Liu, J.; Chen, X.; Pan, F. Altermagnets as a New Class of Functional Materials. *Nat. Rev. Mater.* **2025**, *10* (6), 473–485.

(23) Fender, S. S.; Gonzalez, O.; Bediako, D. K. Altermagnetism: A Chemical Perspective. *J. Am. Chem. Soc.* **2025**, *147* (3), 2257–2274.

(24) Jungwirth, T.; Fernandes, R. M.; Fradkin, E.; MacDonald, A. H.; Sinova, J.; Šmejkal, L. Altermagnetism: An Unconventional Spin-Ordered Phase of Matter. *Newton* **2025**, *1* (6), 100162.

(25) Jungwirth, T.; Sinova, J.; Fernandes, R. M.; Liu, Q.; Watanabe, H.; Murakami, S.; Nakatsuji, S.; Šmejkal, L. Symmetry, Microscopy and Spectroscopy Signatures of Altermagnetism. *Nature* **2026**, *649* (8098), 837–847.

(26) Liu, Y.; Yu, J.; Liu, C.-C. Twisted Magnetic Van Der Waals Bilayers: An Ideal Platform for Altermagnetism. *Phys. Rev. Lett.* **2024**, *133* (20), 206702.

(27) He, R.; Wang, D.; Luo, N.; Zeng, J.; Chen, K.-Q.; Tang, L.-M. Nonrelativistic Spin-Momentum Coupling in Antiferromagnetic Twisted Bilayers. *Phys. Rev. Lett.* **2023**, *130* (4), 046401.

(28) Liu, H.-Z.; He, R.; Zhan, J.-Y.; Wang, D.; He, M.-D.; Luo, N.; Zeng, J.; Chen, K.-Q.; Tang, L.-M. Anomalous Hall Effect in *A*-Type Antiferromagnetic Bilayers. *Phys. Rev. B* **2025**, *112* (13), 134411.

(29) Ruiz, A. M.; Shumilin, A.; González-Hernandez, R.; Baldoví, J. J. Twist-Induced Altermagnetism in a Metallic van Der Waals Antiferromagnet. *arXiv* **2026**, DOI: 10.48550/arXiv.2602.19734.

(30) Dou, M.; Wang, X.; Tao, L. L. Anisotropic Spin-Polarized Conductivity in Collinear Altermagnets. *Phys. Rev. B* **2025**, *111* (22), 224423.

(31) Zhang, X.; Jiang, P.; Xu, L.-Y.; Wang, L.; Liu, L.; Huang, H.-M.; Cao, T.; Li, Y.-L. Giant Spin Splitting and Anisotropic Spin Polarization in 2D Altermagnet $Cr_2O$. *Nano Lett.* **2025**, *25* (46), 16547–16553.

(32) Zarzuela, R.; Jaeschke-Ubiergo, R.; Gomonay, O.; Šmejkal, L.; Sinova, J. Transport Theory and Spin-Transfer Physics in *d*-Wave Altermagnets. *Phys. Rev. B* **2025**, *111* (6), 064422.

(33) Jiang, B.; Hu, M.; Bai, J.; Song, Z.; Mu, C.; Qu, G.; Li, W.; Zhu, W.; Pi, H.; Wei, Z.; Sun, Y.-J.; Huang, Y.; Zheng, X.; Peng, Y.; He, L.; Li, S.; Luo, J.; Li, Z.; Chen, G.; Li, H.; Weng, H.; Qian, T. A Metallic Room-Temperature d-Wave Altermagnet. *Nat. Phys.* **2025**, *21* (5), 754–759.

(34) Asgharpour, A.; Koopmans, B.; Duine, R. A. Synthetic Altermagnets. *Phys. Rev. B* **2025**, *111* (9), 094412.

(35) Zeng, S.; Zhao, Y.-J. Bilayer Stacking *A*-Type Altermagnet: A General Approach to Generating Two-Dimensional Altermagnetism. *Phys. Rev. B* **2024**, *110* (17), 174410.

(36) Boix-Constant, C.; Jenkins, S.; Rama-Eiroa, R.; Santos, E. J. G.; Mañas-Valero, S.; Coronado, E. Multistep Magnetization Switching in Orthogonally Twisted Ferromagnetic Monolayers. *Nat. Mater.* **2024**, *23* (2), 212–218.

(37) Healey, A. J.; Tan, C.; Gross, B.; Scholten, S. C.; Xing, K.; Chica, D. G.; Johnson, B. C.; Poggio, M.; Ziebel, M. E.; Roy, X.; Tetienne, J.-P.; Broadway, D. A. Imaging Magnetic Switching in Orthogonally Twisted Stacks of a van Der Waals Antiferromagnet. *ACS Nano* **2025**, *19* (50), 42140–42147.

(38) Ye, C.; Wang, C.; Wu, Q.; Liu, S.; Zhou, J.; Wang, G.; Söll, A.; Sofer, Z.; Yue, M.; Liu, X.; Tian, M.; Xiong, Q.; Ji, W.; Renshaw Wang, X. Layer-Dependent Interlayer Antiferromagnetic Spin Reorientation in Air-Stable Semiconductor CrSBr. *ACS Nano* **2022**, *16* (8), 11876–11883.

(39) Henríquez-Guerra, E.; Ruiz, A. M.; Galbiati, M.; Cortés-Flores, Á.; Brown, D.; Zamora-Amo, E.; Almonte, L.; Shumilin, A.; Salvador-Sánchez, J.; Pérez-Rodríguez, A.; Orue, I.; Cantarero, A.; Castellanos-Gomez, A.; Mompeán, F.; Garcia-Hernandez, M.; Navarro-Moratalla, E.; Diez, E.; Amado, M.; Baldoví, J. J.; Calvo, M. R. Strain Engineering of Magnetoresistance and Magnetic Anisotropy in CrSBr. *Adv. Mater.* **2025**, 202506695.

(40) Wang, Y.; Luo, N.; Zeng, J.; Tang, L.-M.; Chen, K.-Q. Magnetic Anisotropy and Electric Field Induced Magnetic Phase Transition in the van Der Waals Antiferromagnet CrSBr. *Phys. Rev. B* **2023**, *108* (5), 054401.

(41) Peng, Y.; Ding, S.; Cheng, M.; Hu, Q.; Yang, J.; Wang, F.; Xue, M.; Liu, Z.; Lin, Z.; Avdeev, M.; Hou, Y.; Yang, W.; Zheng, Y.; Yang, J. Magnetic Structure and Metamagnetic Transitions in the van Der Waals Antiferromagnet $CrPS_4$. *Adv. Mater.* **2020**, *32* (28), 2001200.

(42) Ruiz, A. M.; López-Alcalá, D.; Rivero-Carracedo, G.; Shumilin, A.; Baldoví, J. J. Controlling Magnetism in the 2D van Der Waals Antiferromagnet $CrPS_4$ via Ion Intercalation. *Nano Lett.* **2026**, *26* (9), 3018–3025.

(43) Fąs, T.; Wlazło, M.; Birowska, M.; Rybak, M.; Zinkiewicz, M.; Oleschko, L.; Goryca, M.; Gondek, Ł.; Camargo, B.; Szczytko, J.; Budniak, A. K.; Amouyal, Y.; Lifshitz, E.; Suffczyński, J. Direct Optical Probing of the Magnetic Properties of the Layered Antiferromagnet $CrPS_4$. *Adv. Opt. Mater.* **2025**, *13* (10), 202402948.

(44) Lee, J.; Ko, T. Y.; Kim, J. H.; Bark, H.; Kang, B.; Jung, S.-G.; Park, T.; Lee, Z.; Ryu, S.; Lee, C. Structural and Optical Properties of Single- and Few-Layer Magnetic Semiconductor $CrPS_4$. *ACS Nano* **2017**, *11* (11), 10935–10944.

(45) Houmes, M. J. A.; Mañas-Valero, S.; Bermejillo-Seco, A.; Coronado, E.; Steeneken, P. G.; van der Zant, H. S. J. Highly Anisotropic Mechanical Response of the Van Der Waals Magnet $CrPS_4$. *Adv. Funct. Mater.* **2024**, *34* (3), 202310206.

(46) Wu, F.; Gibertini, M.; Watanabe, K.; Taniguchi, T.; Gutiérrez-Lezama, I.; Ubrig, N.; Morpurgo, A. F. Gate-Controlled Magnetotransport and Electrostatic Modulation of Magnetism in 2D Magnetic Semiconductor $CrPS_4$. *Adv. Mater.* **2023**, *35* (30), 2211653.

(47) Yao, F.; Liao, M.; Gibertini, M.; Cheon, C.-Y.; Lin, X.; Wu, F.; Watanabe, K.; Taniguchi, T.; Gutiérrez-Lezama, I.; Morpurgo, A. F. Switching on and off the Spin Polarization of the Conduction Band in Antiferromagnetic Bilayer Transistors. *Nat. Nanotechnol.* **2025**, *20* (5), 609–616.

(48) Chen, Y.; Samanta, K.; Shahed, N. A.; Zhang, H.; Fang, C.; Ernst, A.; Tsymbal, E. Y.; Parkin, S. S. P. Twist-Assisted All-Antiferromagnetic Tunnel Junction in the Atomic Limit. *Nature* **2024**, *632* (8027), 1045–1051.

(49) Peng, Y.; Lin, Z.; Tian, G.; Yang, J.; Zhang, P.; Wang, F.; Gu, P.; Liu, X.; Wang, C.; Avdeev, M.; Liu, F.; Zhou, D.; Han, R.; Shen, P.; Yang, W.; Liu, S.; Ye, Y.; Yang, J. Controlling Spin Orientation and Metamagnetic Transitions in Anisotropic van Der Waals Antiferromagnet $CrPS_4$ by Hydrostatic Pressure. *Adv. Funct. Mater.* **2022**, *32* (7), 202106592.

(50) Telford, E. J.; Chica, D. G.; Ziebel, M. E.; Xie, K.; Manganaro, N. S.; Huang, C.; Cox, J.; Dismukes, A. H.; Zhu, X.; Walsh, J. P. S.; Cao, T.; Dean, C. R.; Roy, X. Designing Magnetic Properties in CrSBr through Hydrostatic Pressure and Ligand Substitution. *Advanced Physics Research* **2023**, *2* (11), 202300036.

(51) Pawbake, A.; Pelini, T.; Mohelsky, I.; Jana, D.; Breslavetz, I.; Cho, C.-W.; Orlita, M.; Potemski, M.; Measson, M.-A.; Wilson, N. P.; Mosina, K.; Soll, A.; Sofer, Z.; Piot, B. A.; Zhitomirsky, M. E.; Faugeras, C. Magneto-Optical Sensing of the Pressure Driven Magnetic Ground States in Bulk CrSBr. *Nano Lett.* **2023**, *23* (20), 9587–9593.

(52) Esteras, D. L.; Rybakov, A.; Ruiz, A. M.; Baldoví, J. J. Magnon Straintronics in the 2D van Der Waals Ferromagnet CrSBr from First-Principles. *Nano Lett.* **2022**, *22* (21), 8771–8778.

(53) Torelli, D.; Thygesen, K. S.; Olsen, T. High Throughput Computational Screening for 2D Ferromagnetic Materials: The Critical Role of Anisotropy and Local Correlations. *2d Mater.* **2019**, *6* (4), 045018.

(54) Olsen, T. Magnetic Anisotropy and Exchange Interactions of Two-Dimensional $FePS_3$, $NiPS_3$ and $MnPS_3$ from First Principles Calculations. *J. Phys. D Appl. Phys.* **2021**, *54* (31), 314001.

(55) Qiu, D. Y.; da Jornada, F. H.; Louie, S. G. Environmental Screening Effects in 2D Materials: Renormalization of the Bandgap, Electronic Structure, and Optical Spectra of Few-Layer Black Phosphorus. *Nano Lett.* **2017**, *17* (8), 4706–4712.

(56) Raja, A.; Chaves, A.; Yu, J.; Arefe, G.; Hill, H. M.; Rigosi, A. F.; Berkelbach, T. C.; Nagler, P.; Schüller, C.; Korn, T.; Nuckolls, C.; Hone, J.; Brus, L. E.; Heinz, T. F.; Reichman, D. R.; Chernikov, A. Coulomb Engineering of the Bandgap and Excitons in Two-Dimensional Materials. *Nat. Commun.* **2017**, *8* (1), 15251.

(57) Lyu, P.; Sødequist, J.; Sheng, X.; Qiu, Z.; Tadich, A.; Li, Q.; Edmonds, M. T.; Zhao, M.; Redondo, J.; Švec, M.; Song, P.; Olsen, T.; Lu, J. Gate-Tunable Renormalization of Spin-Correlated Flat-Band States and Bandgap in a 2D Magnetic Insulator. *ACS Nano* **2023**, *17* (16), 15441–15448.

(58) Lai, J.; Yu, T.; Liu, P.; Liu, L.; Xing, G.; Chen, X.-Q.; Sun, Y. *d*-Wave Flat Fermi Surface in Altermagnets Enables Maximum Charge-to-Spin Conversion. *Phys. Rev. Lett.* **2025**, *135* (25), 256702.

(59) González-Hernández, R.; Šmejkal, L.; Výborný, K.; Yahagi, Y.; Sinova, J.; Jungwirth, T.; Železný, J. Efficient Electrical Spin Splitter Based on Nonrelativistic Collinear Antiferromagnetism. *Phys. Rev. Lett.* **2021**, *126* (12), 127701.

(60) Cui, Q.; Zhu, Y.; Yao, X.; Cui, P.; Yang, H. Giant Spin-Hall and Tunneling Magnetoresistance Effects Based on a Two-Dimensional Nonrelativistic Antiferromagnetic Metal. *Phys. Rev. B* **2023**, *108* (2), 024410.

(61) Železný, J.; Zhang, Y.; Felser, C.; Yan, B. Spin-Polarized Current in Noncollinear Antiferromagnets. *Phys. Rev. Lett.* **2017**, *119* (18), 187204.

(62) Wu, F.; Gibertini, M.; Watanabe, K.; Taniguchi, T.; Gutiérrez-Lezama, I.; Ubrig, N.; Morpurgo, A. F. Magnetism-Induced Band-Edge Shift as the Mechanism for Magnetoconductance in $CrPS_4$ Transistors. *Nano Lett.* **2023**, *23* (17), 8140–8145.

(63) Šmejkal, L.; Hellenes, A. B.; González-Hernández, R.; Sinova, J.; Jungwirth, T. Giant and Tunneling Magnetoresistance in Unconventional Collinear Antiferromagnets with Nonrelativistic Spin-Momentum Coupling. *Phys. Rev. X* **2022**, *12* (1), 011028.

(64) Kresse, G.; Furthmüller, J. Efficient Iterative Schemes for *Ab Initio* Total-Energy Calculations Using a Plane-Wave Basis Set. *Phys. Rev. B* **1996**, *54* (16), 11169–11186.

(65) Mostofi, A. A.; Yates, J. R.; Lee, Y.-S.; Souza, I.; Vanderbilt, D.; Marzari, N. Wannier90: A Tool for Obtaining Maximally-Localised Wannier Functions. *Comput. Phys. Commun.* **2008**, *178* (9), 685–699.